\documentstyle[preprint,aps,epsf]{revtex}
\def\lessim{\mathrel {\vcenter {\baselineskip 0pt \kern 0pt
\hbox{$<$} \kern 0pt \hbox{$\sim$} }}}
\def\gessim{\mathrel {\vcenter {\baselineskip 0pt \kern 0pt
\hbox{$>$} \kern 0pt \hbox{$\sim$} }}}
\def \rightdownarrow
 {\kern.4em {\raise1.75ex\hbox{$\Bigl |$}$\kern-0.27em{\longrightarrow}$}}
\def \mrightdownarrow
 {\kern.4em {\raise1.25ex \hbox{$|$} \kern-0.27em{
                                                \longrightarrow}}}

\def\GeVc{GeV$\!/c$}
\def\GeVcc{GeV$\!/c^2$}
\def\mGeVcc{{\rm GeV\!/c^2}}

\def\mMEt{\not\kern-.35em {E_T}}
\def\MEt{\hbox{$\mMEt$}}

\def\etal{{\it et al.}}
\def\MWprime{M_{W^\prime}}
\font\eightit=cmti8
\def\r#1{\ignorespaces $^{#1}$}
\begin{document}
\title{
\begin{flushright}
FERMILAB-PUB-99/249-E \\
{
\small
CDF/PUB/EXOTIC/CDFR/4841 \\
September 16, 1999 
}
\end{flushright}
%
%
%
Search for a $W^\prime$\ Boson via the Decay Mode
$W^\prime\rightarrow\mu\nu_\mu$\ in 1.8 TeV $p\bar{p}$\ Collisions
}
\author{
%
%
\hfilneg
\begin{sloppypar}
\noindent
F.~Abe,\r {17} H.~Akimoto,\r {39}
A.~Akopian,\r {31} M.~G.~Albrow,\r 7 S.~R.~Amendolia,\r {27} 
D.~Amidei,\r {20} J.~Antos,\r {33} S.~Aota,\r {37}
G.~Apollinari,\r {31} T.~Arisawa,\r {39}
% ...
\end{sloppypar}
\vskip .026in
\begin{center}
(CDF Collaboration)
\end{center}
\vskip .026in
\begin{center}
\r 1  {\eightit Argonne National Laboratory, Argonne, Illinois 60439} \\
\r 2  {\eightit Istituto Nazionale di Fisica Nucleare, University of Bologna,
I-40127 Bologna, Italy} \\
\r 3  {\eightit Brandeis University, Waltham, Massachusetts 02254} \\
\r 4  {\eightit University of California at Los Angeles, Los 
Angeles, California  90024} \\  
\r 5  {\eightit University of Chicago, Chicago, Illinois 60637} \\
\r 6  {\eightit Duke University, Durham, North Carolina  27708} \\
\r 7  {\eightit Fermi National Accelerator Laboratory, Batavia, Illinois 
60510} \\
\r 8  {\eightit University of Florida, Gainesville, Florida  32611} \\
\r 9  {\eightit Laboratori Nazionali di Frascati, Istituto Nazionale di Fisica
               Nucleare, I-00044 Frascati, Italy} \\
\r {10} {\eightit University of Geneva, CH-1211 Geneva 4, Switzerland} \\
\r {11} {\eightit Harvard University, Cambridge, Massachusetts 02138} \\
\r {12} {\eightit Hiroshima University, Higashi-Hiroshima 724, Japan} \\
\r {13} {\eightit University of Illinois, Urbana, Illinois 61801} \\
\r {14} {\eightit Institute of Particle Physics, McGill University, Montreal 
H3A 2T8, and University of Toronto,\\ Toronto M5S 1A7, Canada} \\
\r {15} {\eightit The Johns Hopkins University, Baltimore, Maryland 21218} \\
\r {16} {\eightit Institut f\"{u}r Experimentelle Kernphysik, 
Universit\"{a}t Karlsruhe, 76128 Karlsruhe, Germany} \\
\r {17} {\eightit National Laboratory for High Energy Physics (KEK), Tsukuba, 
Ibaraki 305, Japan} \\
\r {18} {\eightit Ernest Orlando Lawrence Berkeley National Laboratory, 
Berkeley, California 94720} \\
\r {19} {\eightit Massachusetts Institute of Technology, Cambridge,
Massachusetts  02139} \\   
\r {20} {\eightit University of Michigan, Ann Arbor, Michigan 48109} \\
\r {21} {\eightit Michigan State University, East Lansing, Michigan  48824} \\
\r {22} {\eightit University of New Mexico, Albuquerque, New Mexico 87131} \\
\r {23} {\eightit The Ohio State University, Columbus, Ohio  43210} \\
\r {24} {\eightit Osaka City University, Osaka 588, Japan} \\
\r {25} {\eightit Universita di Padova, Istituto Nazionale di Fisica 
          Nucleare, Sezione di Padova, I-35131 Padova, Italy} \\
\r {26} {\eightit University of Pennsylvania, Philadelphia, 
        Pennsylvania 19104} \\   
\r {27} {\eightit Istituto Nazionale di Fisica Nucleare, University and Scuola
               Normale Superiore of Pisa, I-56100 Pisa, Italy} \\
\r {28} {\eightit University of Pittsburgh, Pittsburgh, Pennsylvania 15260} \\
\r {29} {\eightit Purdue University, West Lafayette, Indiana 47907} \\
\r {30} {\eightit University of Rochester, Rochester, New York 14627} \\
\r {31} {\eightit Rockefeller University, New York, New York 10021} \\
\r {32} {\eightit Rutgers University, Piscataway, New Jersey 08855} \\
\r {33} {\eightit Academia Sinica, Taipei, Taiwan 11530, Republic of China} \\
\r {34} {\eightit Texas A\&M University, College Station, Texas 77843} \\
\r {35} {\eightit Texas Tech University, Lubbock, Texas 79409} \\
\r {36} {\eightit Istituto Nazionale di Fisica Nucleare, University of Trieste/
Udine, Italy} \\
\r {37} {\eightit University of Tsukuba, Tsukuba, Ibaraki 305, Japan} \\
\r {38} {\eightit Tufts University, Medford, Massachusetts 02155} \\
\r {39} {\eightit Waseda University, Tokyo 169, Japan} \\
\r {40} {\eightit University of Wisconsin, Madison, Wisconsin 53706} \\
\r {41} {\eightit Yale University, New Haven, Connecticut 06520} \\
\end{center}
}
\draft
\address{}
\maketitle
%
%

\begin{abstract}
We report the results of a search for a $W^\prime$\ boson produced 
in $p\bar p$\ collisions  at a center-of-mass energy of 1.8 TeV
using a $107~\hbox{pb}^{-1}$\ data sample recorded by the Collider Detector at
Fermilab. We consider the decay channel $W^\prime\rightarrow
\mu\nu_\mu$\ and search for anomalous production of high transverse mass
$\mu\nu_\mu$\ lepton pairs.  We observe no excess of events above 
background and set limits on the rate of $W^\prime$\ boson production and decay
relative to Standard Model $W$\ boson production and decay
using a fit of the transverse mass distribution observed.
If we assume Standard Model strength couplings of
the 
$W^\prime$\ boson to quark and lepton pairs, we exclude a $W^\prime$\
boson with invariant mass less than $660~\hbox{GeV}\!/c^2$\ at 95\%
confidence level.
\end{abstract}

%
%

%
Three of the four known forces of nature, the strong, electromagnetic
and weak force, are described by the Standard Model using a local gauge
theory that accounts for each interaction using a vector boson force
carrier \cite{ref: EWK Reference}.  
The predictions of this model have been confirmed by the discoveries
of the $W$\ and $Z^0$\
bosons, the carriers of the weak force, and high precision
measurements of their properties.
The Standard Model is not
a complete theory, however, as it fails to explain the number of lepton
and quark generations, the rather large mass scale between the very
lightest and very heaviest of the fundamental fermions, and 
the number or structure of the gauge symmetries that exist in nature.
It is still an open experimental question as to whether additional
forces exist.   Evidence for a new force could come from observation of the 
corresponding force carrier.

Previous searches have been conducted for
possible new force carriers that couple to $\mu$\ final states in a manner similar
to the vector bosons that mediate the weak force.
These searches have yielded null results, and have set 
model-dependent limits on the rate at which such a particle is produced and its
mass.  The most sensitive searches have been performed at the Fermilab Tevatron
Collider.  A
$Z^\prime$\ boson with a mass $<690~\hbox{GeV}\!/c^2$\ has been excluded at 95\%
confidence level (CL) \cite{ref: Zprime searches}. 
Searches considering the decay mode
$W^\prime\rightarrow\mu\nu_\mu$\ have excluded a $W^\prime$\ boson with
mass $<435~\hbox{GeV}\!/c^2$\ at 95\%\ CL \cite{ref: Wprime searches muon}. These
mass limits all assume that the new vector boson's couplings to leptonic
final states will be given by the Standard Model, which
predicts that the total width of the boson increases linearly with $\MWprime$,
where $\MWprime$\ is the mass of the boson.
Indirect searches studying, for example, the Michel spectrum in $\mu$\ decay
have resulted in more model-independent limits with less 
sensitivity \cite{ref: indirect searches}.
Searches in other channels have also been used to place constraints on possible
$W^\prime$\ masses: The most stringent exclude a
$W^\prime$\ boson at 95\%\ CL with a mass
$<720~\hbox{GeV}\!/c^2$\ that decays via $W^\prime\rightarrow e\nu_e$\
\cite{ref: Wprime searches electron}.  

In this letter, we present the results
of a new search for a
$W^\prime$\ boson in the $\mu\nu_\mu$\ decay mode.  We use a data
sample of $107~\hbox{pb}^{-1}$\ of 1.8 TeV $p\bar p$\ collisions recorded by the
Collider Detector at Fermilab (CDF) detector during 1992-95.  This search
is based on an analysis of high mass $\mu\nu_\mu$\ candidate final states,
and is sensitive to a variety of new phenomena that would result in 
anomolous production of such high mass events.
We use these data to set limits
on the production cross section times branching fraction of the process 
\begin{eqnarray}
p\bar{p} \rightarrow W^\prime X \rightarrow \mu\nu_\mu X,
\end{eqnarray}
normalizing the candidate event sample to the
large observed $W\rightarrow\mu\nu_\mu$\ signal in the same event sample.
This search assumes that the decay $W^\prime\to WZ^0$\ is suppressed
\cite{ref: WprimetoWZ} and that $M_\nu \ll M_{W^\prime}$, where $M_\nu$
is the mass of the neutrino from a $W^\prime$\ boson decay.
We also assume that the daughter neutrino does not decay within
the detector volume.

In this search, we select events that are consistent with the
production of both the Standard Model $W$\ boson, followed by the decay
$W\rightarrow\mu\nu_\mu$, and any heavier object that decays in
the same manner.  We place limits on the production and decay rate of
such a massive object relative to the production and decay rate of the
$W$\ boson.   This approach avoids the need to make an absolute cross
section measurement or upper limit, and avoids many of the uncertainties
associated with such a technique.  We subsequently use our relative
production and decay rate upper limits to set lower limits on the mass of
such a
$W^\prime$\ boson. However, these upper limits place constraints on any
processes that generate high mass $\mu\nu_\mu$\ pairs, and represent an
increase of a factor of 20 in sensitivity from earlier searches
in this channel.
Additional details of this analysis are presented in Ref.
\cite{ref: thesis}.

The CDF detector is described in detail elsewhere
\cite{ref: CDF Detector}. 
The detector has a charged particle tracking
system immersed in a 1.41~T solenoidal magnetic field, which is coaxial
with the $p\bar p$\ beams. 
The tracking system consists of solid state tracking
detectors and drift chambers that measure particle momentum with an
accuracy of 
$
\sigma_{p_T}/p_T \sim 0.001 p_T,
$
where $p_T$\ is the momentum of the charged particle measured in \GeVc\
transverse to the $p\bar p$\ beam line.
The tracking system is surrounded by segmented electromagnetic and
hadronic calorimeters that measure the flow of energy associated with
particles that interact hadronically or electromagnetically out to a
pseudorapidity
$|\eta|$\ of 4.2~\cite{ref: Coord-system}.
A set of charged particle detectors
outside the calorimeter is used to identify muon candidates
with $|\eta|<1.0$.

Candidate events were identified in the CDF
trigger system by the requirement of at least one muon candidate with
$p_T>9$\ or 12~\GeVc, depending on running conditions.  
The event sample was subsequently
refined after full event reconstruction by requiring a well-identified
muon candidate with momentum $p_T>20$~\GeVc\ and by requiring that the
missing transverse energy in the event, $\MEt$, be greater than 20~GeV. 

Additional requirements were
imposed to reject specific sources of backgrounds.   Events consistent
with arising from QCD dijet production, where one jet is misidentified as
a muon candidate, were rejected by requiring that the muon candidate was
isolated from energy flow in the event and that the energy deposited in
the calorimeter by the muon candidate was consistent with that arising
from a minimum ionizing particle.   Events 
due to Drell-Yan production of dimuons (dominated
by the decay
$Z^0\rightarrow
\mu^+\mu^-$) were rejected by vetoing events if a second
isolated muon candidate with $p_T>15$~\GeVc\ was found in the event.
Finally, events arising from cosmic rays
were rejected by imposing tight requirements between the
timing of the beam interaction and the muon candidate passing through
the calorimeter, and by removing events that had evidence of a second
charged particle observed within 0.05 radians of being back-to-back with the
$\mu$\ candidate.

\begin{figure}
\begin{center}
\leavevmode
\hbox{%
\epsfxsize=5.4in
\epsffile{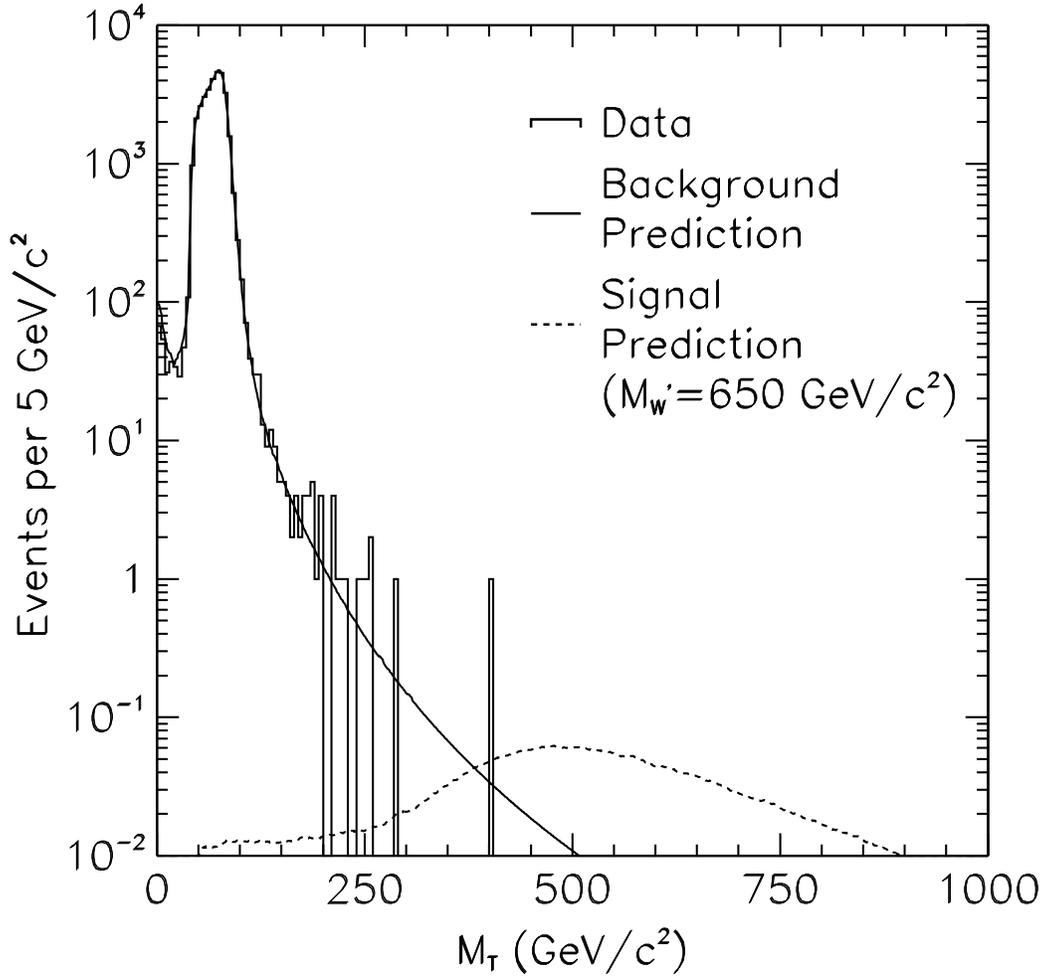}
}
\end{center}
\caption
{The transverse mass spectrum of the $\mu\nu_\mu$\ candidate events.
The background rate is predicted from the fit described in the text.  The 
distribution expected from the production of a 
$W^\prime$\ boson with a mass of 650~\GeVcc\ is illustrated by the dashed
distribution.}
\label{fig: MT distribution}
\end{figure}

This selection resulted in a sample of 31\,992 events.  The
distribution of the transverse mass
\begin{eqnarray}
M_T \equiv \sqrt{2 p_T \MEt (1-\cos\phi_{\mu\nu})},
\end{eqnarray}
where $\phi_{\mu\nu}$\ is the azimuthal angle between the $\mu$\
candidate and the missing transverse energy vector, shows a clear
Jacobian peak that is associated with the production and decay of the
$W$~boson.  This distribution, illustrated in 
Fig.~\ref{fig: MT distribution}, 
also shows a smoothly falling distribution above the Jacobian
peak with little obvious structure.  

In order to understand the composition of this high transverse mass
sample, we fit the $M_T$\ distribution between 40 and 2000~\GeVcc\
using an unbinned maximum
likelihood technique, which included contributions from a
hypothetical
$W^\prime$\ boson decaying to the $\mu\nu_\mu$\ final state, 
$W\rightarrow\mu\nu_\mu$\ decay and all other significant background sources.  
The largest background sources were 
the production and decay of the $W$\ and
$Z^0$\ bosons into final states consisting of muons.  These included the
decay modes
$W\rightarrow\mu\nu_\mu$, $W\rightarrow\tau\nu_\tau\rightarrow
\mu\nu_\mu\nu_\tau$, $Z^0\rightarrow\tau^+\tau^-\rightarrow \mu X$,
and $Z^0\rightarrow\mu^+\mu^-$.  The other background sources were
muons arising from top quark production and ``fake'' muons arising from
QCD dijet production.
The shape of the $M_T$\ distributions for the
$W^\prime$\ signal and the backgrounds from $W$\ and
$Z^0$\ production were calculated using a Monte Carlo
technique employing the PYTHIA programme 
\cite{ref: PYTHIA}.
We used a next-to-leading order theoretical
prediction for the
$p_T$\ and $\eta$\ dependence of $W^\prime$\ and $W$\ production
\cite{ref: Pt and eta dependence}.  
Our model included a simulation of the CDF detector that was derived from 
studies of $Z^0\rightarrow\mu^+\mu^-$\ candidate events.

Studies of specific data samples constrained the size and
shape of the other possible background contributions.  
The relative size of the various
$W$\ and $Z^0$\ boson decay modes and $t\bar{t}$\ production
were determined using the measured
production ratios and 
branching fractions to these final states \cite{ref: PDG}.  The size
of the dijet background was determined by studying the characteristics
of event samples enriched in this dijet contamination.
The total number of events with 
$M_T>200~\hbox{GeV}/c^2$\ from Standard Model sources was estimated 
from the fit to the $M_T$\ distribution between 40 and 2000~\GeVcc\ to be
$11.8\pm0.9$\ events, with the largest contribution arising from 
off-mass-shell $W$\ boson production. This agrees with the 
observed yield of 14 events in this $M_T$\ region. 

The results of the fit to the $M_T$\ data distribution assuming only
contributions from $W$\ production and decay and the other known
background sources are plotted in  Fig.~\ref{fig: MT distribution}. The
agreement between the data distribution and the fit prediction is good. 
A small excess of events with transverse masses around 200~\GeVcc\ is not
statistically significant.
The contributions 
from the various
background sources are listed in Table~\ref{tab: event rates}.

Our Monte Carlo calculation together with the
detector model was used to determine the ratio of acceptances for detection of
$W^\prime$\ and $W$\ bosons.  This ratio rises 
as a function of $M_{W^\prime}$, peaking at $\sim1.7$\ for 
$M_{W^\prime}=300~\hbox{GeV}\!/c^2$, and then falling to $\sim1.5$\ for
$\MWprime=800~\hbox{GeV}\!/c^2$. The initial increase in
acceptance is due to a heavier $W^\prime$ boson being
produced more centrally.
The subsequent decrease results from very
high energy muon daughters depositing significant amounts of energy in the
calorimeter.

We set upper limits on the relative contribution of a
$W^\prime$\ boson by fitting the data distribution to a
combination of the background distributions described above and a
$W^\prime$\ $M_T$\ distribution
expected from the production and decay of a $W^\prime$\ boson of
a given mass.  
The results of the fit, expressed as the ratio of observed $W^\prime$ boson
candidates to the number expected assuming Standard Model 
strength couplings, are shown in Table~\ref{tab: fit results}.
We then used the resulting likelihood function
to set a 95\%\ CL upper limit on this ratio, 
also shown in Table~\ref{tab: fit results}.  In setting these limits, we only
considered the likelihood function in the ``physical region'' where this ratio
was greater than or equal to zero.  We note that these limits are insensitive to
the assumed width of the $W^\prime$\ boson, as the width of the expected signal
distribution is dominated by detector resolution for $W^\prime$\ masses greater
than approximately 300~\GeVcc.

The procedure used to calculate this upper limit
incorporated various systematic uncertainties using the method given in
\cite{ref: PDG}.  The largest 
resulted from the choice of parton distribution function, which at the
highest masses contributed
$\sim\pm10\%$\ uncertainty to the relative $W$\ and $W^\prime$
production cross section.  We used the CTEQ~4A1 parton distribution
functions with a four-momentum transfer squared $Q^2 = \MWprime^2$\ for
our result \cite{ref: CTEQ4A1} but employed several parton distribution
functions to determine our sensitivity to this choice. Other systematic
uncertainties included those arising from our knowledge of the track
$p_T$\ resolution 
and the uncertainty in acceptance
arising from variations in the
$W^\prime$\ boson 
$p_T$\ distribution.
The total
systematic uncertainty 
varied from 4\% for
$\MWprime=200~\hbox{GeV}\!/c^2$\ 
to 12\% for $\MWprime=700$~\GeVcc. These were
incorporated into our upper limits using a procedure that convoluted
the likelihood function determined by our fit to the $M_T$\
distribution with the probability distribution functions associated
with each uncertainty. The results are dominated by the statistical
uncertainties of the data sample.  We also computed cross section upper limits by
counting signal events above background in the high transverse mass region and
obtained comparable results to the likelihood fit, though these depended on the
region chosen for signal events.

We can convert the 95\%\ CL upper limit on the relative cross sections and
decay rates into a lower limit on the mass of the $W^\prime$\ boson by excluding
all masses where our 95\%\ CL limit on the ratio of cross sections times
branching fractions is less than unity.  We determined the
predicted cross sections using a parton-level matrix 
element calculation and the CTEQ~4A1 parton distribution functions, 
taking into account the
fact that a $W^\prime$\ boson with a mass above approximately
$180~\hbox{GeV}\!/c^2$\ decays into three quark generations.

\begin{figure}
\begin{center}
\leavevmode
\hbox{%
\epsfxsize=5.4in
\epsffile{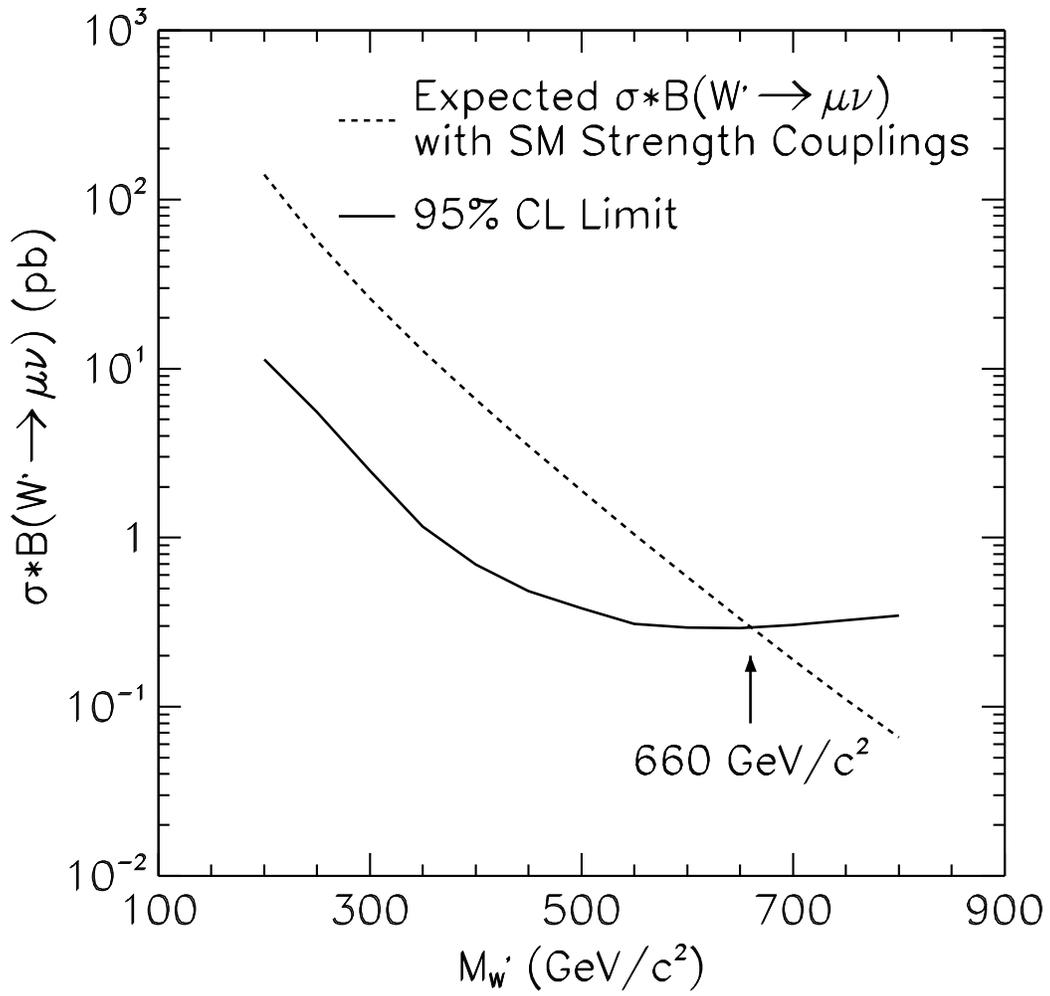}
}
\end{center}
\caption{
The upper limits on the $W^\prime$\ boson production cross section as a
function of the $W^\prime$\ boson mass. }
\label{fig: upper limits}
\end{figure}

The resulting upper limit on the $W^\prime$\ boson cross section versus
$M_{W^\prime}$\ is shown in Fig.~\ref{fig: upper limits}, where we
have now normalized our upper limits on the production cross section
ratios using the predicted
$W$\ boson production cross section,
which is consistent with measurements \cite{ref: W boson sigma}.  
We compare
this upper limit with the predictions for a $W^\prime$\ boson with Standard
Model strength couplings, also shown in the figure.
This allows us to exclude a $W^\prime$
boson with mass between 200 and $660~\hbox{GeV}\!/c^2$.
Taking into account the previous searches in this channel, 
a $W^\prime$ boson with Standard Model strength couplings
and mass below $660~\hbox{GeV}\!/c^2$\ can be excluded.  This corresponds to an
increase in sensitivity of approximately a factor of 20 from the earlier studies
of this final state.

In summary, we have performed a search for the production of a new
heavy vector gauge boson in 1.8 TeV $p\bar{p}$\ collisions and decaying
into the $\mu\nu_\mu$\ final state.  
We use a fit of the $M_T$\ distribution to exclude a
$W^\prime$\ boson with mass $<660~\hbox{GeV}\!/c^2$\ at 95\%\ CL, assuming
Standard Model strength couplings.  This limit is comparable to 
those set using the $e\nu_e$\ decay modes, and represents a significant
improvement in sensitivity for $W^\prime$\ boson searches using the muon
decay mode.

\section*{Acknowledgements}
We thank the Fermilab staff and the technical staff at the
participating institutions for their essential contributions to this
research.  This work is supported by the U.~S.~Department of Energy
and the National Science Foundation; the Natural Sciences and
Engineering Research Council of Canada; the Istituto Nazionale di
Fisica Nucleare of Italy; the Ministry of Education, Science and
Culture of Japan; the National Science Council of the Republic of
China; and the A.~P.~Sloan Foundation.

%

%
%

\newpage


%
\begin{table}
\caption{The event yields for the background sources 
above and below $M_T=200$~\GeVcc.
Uncertainties are correlated.}
\begin{center}
\begin{tabular}{lcc}
Process       &   \multicolumn{2}{c}{Fitted Event Yield}  \\
     &   $(40< M_T < 200~\mGeVcc)$   &  $(M_T>200~\mGeVcc)$ \\
\hline
$W\rightarrow\mu\nu_\mu$         &$27\,925\pm 209$  &$8.99\pm 0.81$ \\
$W\rightarrow\tau\nu_\tau$       &$\    687\pm  27$  &$0.04\pm 0.01$ \\
$Z/\gamma\rightarrow\mu\mu$      &$\ 2\,824\pm 196$  &$2.02\pm 0.35$ \\
$Z/\gamma\rightarrow\tau\tau$    &$\     47\pm   3$  &$0.02\pm 0.02$ \\
$t\bar{t}$\                      &$\  14^{+4}_{-3}$  &$0.29^{+0.07}_{-0.06}$ \\
QCD                              &$\     74\pm  37$  &$0.42^{+0.43}_{-0.42}$ \\
\end{tabular}
\end{center}        
\label{tab: event rates}
\end{table}

%
%
%
\begin{table}
\caption{
The expected number of 
events from $W^\prime$\ boson production, 
$N_{exp}$, assuming Standard Model strength couplings and normalized to the
observed $W$~boson yield.
We also show 
the rate of $W^\prime$\ boson production and decay
relative to the rate predicted using
Standard Model couplings, 
and the 95\%\ CL upper limit on this relative rate as a function of
$\MWprime$.
The uncertainties are 
statistical and do not include the systematic uncertainties.
The 95\%\ CL upper limit includes both statistical and systematic
uncertainties.
}
\begin{center}
\begin{tabular}{cccc}
$M_{W^\prime}$~(\GeVcc)
&$N_{exp}$~(Events)      
&\multicolumn{2}{c}{${\sigma\cdot{\cal B}(W^\prime\rightarrow
\mu\nu_\mu)}\over {\sigma\cdot{\cal B}(W^\prime\rightarrow
\mu\nu_\mu)_{SM}}$} \\
& &Fit &Upper Limit
\\
\hline
200   &   $2330\pm 100$  & $0.009^{+0.004}_{-0.004}$ & 0.08 \\
250   &   $984\pm 45$  & $0.011^{+0.007}_{-0.006}$ & 0.10 \\
300   &   $456\pm 26$  & $0.006^{+0.011}_{-0.006}$ & 0.10 \\
350   &    $224\pm 13$  & $0.000^{+0.014}_{-0.006}$ & 0.09 \\
400   &    $115\pm ~8$ & $0.000^{+0.018}_{-0.002}$ & 0.11 \\
450   &    $60.2\pm 3.5$ & $0.000^{+0.026}_{-0.000}$ & 0.14 \\
500   &    $32.5\pm 2.4$ & $0.000^{+0.039}_{-0.000}$ & 0.20 \\
550   &    $17.2\pm 1.4$ & $0.000^{+0.058}_{-0.000}$ & 0.30 \\
600   &    $9.69\pm 0.84$ & $0.000^{+0.096}_{-0.000}$ & 0.50 \\
650   &   $5.37\pm 0.50$  & $0.000^{+0.159}_{-0.000}$ & 0.83 \\ 
700   &     $3.01\pm 0.31$ & $0.000^{+0.273}_{-0.000}$ & 1.90 \\
750   &     $1.72\pm 0.21$ & $0.000^{+0.495}_{-0.000}$ & 2.94 \\
\end{tabular}
\end{center}
\label{tab: fit results}
\end{table}


\begin{references}
\bibitem{ref: Coord-system}
We use a coordinate system where the polar
angle to the proton beam is denoted by $\theta$, $\phi$\ is the
azimuthal angle about this beam axis, and
$\eta\equiv-\ln\tan(\theta/2)$.
\end{references}
\end{document}